\documentclass[3p,11pt]{elsarticle}
\usepackage{amsthm}           
\usepackage[utf8]{inputenc}
\usepackage{caption}
\usepackage{subcaption}
\usepackage{graphicx}
\usepackage{xcolor}
\usepackage{placeins}
\usepackage{amssymb}
\usepackage{mathtools}
\usepackage{epstopdf}
\usepackage{varioref}

\renewcommand{\vec}[1]{\boldsymbol{#1}}
\newcommand{\diff}{\ensuremath{\mathrm d}}
\newtheorem*{theorem}{Theorem}
\newtheorem*{definition}{Definition}

\begin{document}

\begin{frontmatter}
\title{ About the role of chaos and coarse graining in Statistical Mechanics}
\author{G. Falasco, G. Saggiorato, A. Vulpiani}

\begin{abstract}
We discuss  the role of ergodicity and chaos for the validity of
statistical laws. 
In particular we explore   the basic aspects of chaotic systems
(with emphasis  on the finite-resolution) 
on systems composed of a huge number of particles.
\end{abstract}
\end{frontmatter}

\section{Introduction}

Statistical mechanics was founded by Maxwell, Boltzmann and Gibbs to account for
the thermodynamics of  macroscopic bodies,  i.e. systems with a very large number of particles,
starting from the microscopic interactions.
The typical claim is   that statistical mechanics works
without very precise requirements on the underlying  dynamics (except for the assumption of
ergodicity).
With the discovery of deterministic chaos it became clear that statistical
approaches may also be unavoidable and useful
in systems with few degrees of freedom. However, even after many years there is
no general agreement among the experts about the fundamental ingredients for the
validity of statistical mechanics.

The wide spectrum of positions
ranges  from the belief of Landau and Khinchin in the main role of the many
degrees of freedom and the (almost) complete irrelevance of dynamical properties,
in particular ergodicity, to the opinion of those, for example Prigogine and his
school, who consider chaos as the basic ingredient.

For almost all practical purposes one can say that the whole subject of statistical
mechanics consists in the evaluation of a few suitable quantities (for example, the
partition function, free energy, correlation functions). The ergodic problem is often
forgotten and the (so-called) Gibbs approach is accepted because ``it works''.
 Such
a point of view cannot be satisfactory, at least if one believes that it is not less
important to understand the foundation of such a complex issue than to calculate
useful quantities.

The present paper  is meant to serve as a short  informal introduction to the problem
of the connection between dynamical behavior (mainly ergodicity and chaos) and
statistical laws.

This contribution has been written by G.S. and G.F from the lectures delivered by A.V.

\section{The bridge to statistical mechanics}
Before moving on to a more technical discussion we want to elucidate the pivotal role played by ergodicity in passing from a mechanical to a statistical description of a macroscopic system. Let us consider an isolated system of  $N$ particles. Denoting by $\vec{q}_i$ and $\vec{p}_i$ the position and momentum vectors of the $i$th particle, the state of the system at time $t$ is described by the vector $\vec x(t)=(\vec{q}_1(t),...,\vec{q}_N(t),\vec{p}_1(t),...,\vec{p}_N(t))$ in a $6N$-dimensional phase space. The evolution law is given by Hamilton's equations. Given a physical observable $A(\vec{x})$, its time average over a period $T$ is defined by:
\begin{equation}\label{Tave}
\overline{A}^T(\vec{x}(0))= \frac{1}{T}\int_{0}^{T}A(\vec{x}(t))\text{d}t\,.
\end{equation}
In general \eqref{Tave} is a function of the averaging time $T$ and of the initial conditions $\vec{x}(0)$. We assume that $\overline{A}^T$ is the result of an experimental measurement, i.e. is performed on a macroscopic time $T$. Therefore, since $T$ is much larger than the microscopic dynamics time scale over which the molecular changes occur, the formal limit $T \to \infty$ can be considered. We say that the ergodic hypothesis is satisfied if the time average 
\begin{equation}\label{Tave2}
\overline{A} \equiv \lim_{T\to \infty} \frac{1}{T}\int_{0}^{T} A(\vec{x}(t))\diff t,
\end{equation}
is a function of the energy only, independent of the initial conditions. If it is the case, the time average \eqref{Tave2} can be replaced by the probability average
\begin{equation}\label{Pave}
\overline{A} = \int \rho_{\text{mc}}(\vec x) A(\vec x) \diff\vec x \equiv  \langle A \rangle,
\end{equation}
where $\rho_{\text{mc}}(\vec x)$ is the micro-canonical probability density function, defined constant over a constant energy surface. The validity of \eqref{Pave} constitutes the main question of the ergodic problem. In the following we address the issue of identifying the conditions under which a dynamical system is ergodic. It will soon be clear that mathematical ergodicity is extremely difficult to be asserted and, in fact, weaker properties can be required to physical systems.

Let us wonder about the basic question of the foundation of the statistical mechanics:
why do the  time averages of the microscopic
properties of a system of particles describe  the thermodynamic properties of
the object composed of those particles?
 The microscopic system, in fact, consists of a myriad
of mechanical parameters, such as the particles, energy and momentum,
while thermodynamics consists of just a few measurable quantities, like temperature and pressure. 
 The corresponding diversity of fundamental terminology,
qualifies the reduction of thermodynamics to mechanics as ``heterogeneous''
reduction \cite{NAGEL1979}, a condition which may prevent the logical derivation
of the former theory from the latter.
 For such a problem  the typical approach of the philosophy of science is to 
 require the existence of relations between the terms
of the mechanics and elements of the vocabulary of thermodynamics.
Such bridge law must reflect a kind of identity between the objects
of study of the two theories \cite{NAGEL1979}.
The bridge law which associates thermodynamics with the classical mechanics
of atoms was proposed by Boltzmann and it is engraved in his tomb
stone:
\begin{equation}
\label{Boltz}
S = k \log W \,.
\end{equation}
 This celebrated relation connects the thermodynamic entropy $S$ of an object
in the macroscopic state $X$, to the volume $W$  of all microstates 
correspond to the same $X$. For example, considering the macrostate $X$
corresponding to a given energy $E$, one typically considers the energy shell
$E - \delta E \le H(\{ {\bf q}_n \} , \{ {\bf p}_n \}) < E$, with small $\delta E$, and obtains:
 $$
 W = \int_{E - \delta E \le H({\bf q}, {\bf p}) < E} 
 d{\bf q}_1\dots d{\bf q}_N  d{\bf p}_1\dots d{\bf p}_N \,\,
 $$
 The micro-canonical probability distribution is constant (equal to $1/W$) in the energy shell
and zero otherwise.
Equation \eqref{Boltz} qualifies as a bridge law, because $S$ is a thermodynamic
quantity, while $W$ is a microscopic entity. Once it has been introduced,
further mechanical properties of our description of the microscopic dynamics
may be related to as many other thermodynamic quantities, thus bridging
the gap between micro- and macro-descriptions.
\\

\section{Some general consideration about ergodicity}
Here we introduce the notion of dynamical system:
\begin{definition}[Dynamical system]\ \\
A deterministic dynamical system is described by the triplet $(\Omega, S^t, \mu)$, where:
\begin{itemize}
\item $\Omega$ is the phase space containing the system state vector $\vec x$;
\item $S^t$ is the evolution operator: $\vec{x}(t)=S^t\vec{x}(0)$;
\item $\mu$ is a measure invariant under the evolution $S^t$, i.e. $\mu(G)=\mu(S^{-t} G)$ for all $G\subset \Omega$.
\end{itemize}
\end{definition}
 The consideration of the previous section naturally extend to a generic dynamical system, leading to the following definition of ergodicity \cite{ARNOLD68}:
\begin{definition}[Ergodicity]\ \\
The dynamical system  $(\Omega, S^t, \mu)$ is called ergodic with respect to the invariant measure $\mu(\vec{x})$, if, for every integrable function $A(\vec x)$ and for almost all initial conditions $\vec{x}(0)$, one has:
\begin{equation}\label{ergodic}
\overline{A}\equiv\lim_{T\to \infty} \frac{1}{T}\int_{0}^{T} A(S^t\vec{x}(0))\diff t = \int A(\vec x)\diff \mu(\vec x) \equiv  \langle A \rangle .
\end{equation}
\end{definition}
  We note that in conservative systems (as the isolated Hamiltonian system of the previous section) the invariant measure can be written in terms of a density, i.e. $\diff \mu(\vec x)= \rho(\vec x) \diff \vec x$, while in dissipative systems it cannot, being singular with respect to the Lebesgue measure due to the shrinking of space phase volumes. \\
On an abstract level, necessary and sufficient conditions for a dynamical system to be ergodic are given by Birkhoff theorem \cite{BIRKHOFF31}:
\begin{theorem}[Birkhoff]\
\begin{itemize}
\item For almost every initial condition $\vec{x}(0)$ the following time average exists:
\begin{equation}\label{Birkhoff}
\overline{A}(\vec{x}(0)) = \lim_{T\to \infty} \frac{1}{T}\int_{0}^{T} A(S^t\vec{x}(0))\diff t\,.
\end{equation}
\item  The system is ergodic if, and only if, $\Omega$ cannot be subdivided into two invariant parts each of positive measure.
\end{itemize}
\end{theorem}
Since in general it is not possible to decide whether a given system satisfies the above hypothesis, the Birkhoff theorem is of scarce practical relevance in connection with statistical mechanics. 
Also, since the limit $T\to \infty$ in \eqref{Birkhoff} is just formal, one should ask oneself how large $T$ has to be taken in practice in order to get a fair convergence of time and ensemble averages. It turns out that, for microscopic observables, $T$ is of the same order of the Poincar\'{e} recurrence time. As a simple example we consider a phase space region $G \subset\Omega$ and the observable
\begin{equation}\label{B}
B(\vec x)= 
\begin{dcases}
1 \;\; \text{if } \vec x \in G,\\
0 \;\; \text{otherwise}.
\end{dcases}
\end{equation}
We call $B(\vec x)$ a microscopic observable because the exit of a single particle from $G$ causes a variation of $B(\vec x)$ of the same order of its value. The time average $\overline{B}^{T}$ is the fraction of time spent by the system in $G$ during the interval $[0,T]$, and it approaches $\mu(G)$ when $T \to \infty$. Clearly, the time needed to have a good agreement of $\overline{B}^{T}$ with $\langle B\rangle $ is of the order of the time, denoted $\langle \tau(G) \rangle$, it takes the system to explore the phase space and to come back to $G$. Thanks to the \emph{Kac lemma} we can express this mean recurrence time as \cite{KAC59}:
\begin{equation*}
\langle \tau(G) \rangle=\frac{\tau_0}{\mu(G)},
\end{equation*}
where $\tau_0$ is a characteristic time of the system. Assuming $G$ to be a $6N$-dimensional hypercube of edge $l_\text{G}$ in a phase space of characteristic dimension $L > l_\text{G}$, we can use the estimation $\mu(G) \sim (l_\text{G}/L)^{6N}$ to obtain:
\begin{equation}\label{Kac}
\langle \tau(G) \rangle \sim \tau_0 \left(\frac{L}{l_\text{G}}\right)^{6N}.
\end{equation}
From \eqref{Kac} we conclude that, in macroscopic systems, time averages of some microscopic quantity must be performed for times $T$ much larger than the age of the Universe in order to be in accordance with ensemble averages. 

Nevertheless, it  must be recognized that the approach we have adopted so far, consisting in proving ergodicity for generic systems and observables, is quite demanding and in fact not useful for the purpose of justifying the statistical approach. In this regard, it is more reasonable to require ergodicity to hold in the weaker form
\begin{equation}\label{weak}
\overline{A}=\langle A \rangle + O(\epsilon) \;\;\; \text{ with } \epsilon \ll1,
\end{equation} 
after having introduced the following assumptions:
\begin{itemize}
\item the system is macroscopic, i.e composed of $N\gg1$ particles;
\item equation \eqref{weak} is valid for physical observables, not for all generic functions.
\item  equation \eqref{weak} is valid for initial conditions $\vec{x}(0)\in \Omega\!-\!G$, with $\mu(G) \ll1$. 
\end{itemize}
The validity of \eqref{weak} for specific macroscopic observables and a particular class of Hamiltonian systems is ensured by the Khinchin theorem \cite{KHINCHIN57}:
\begin{theorem}[Khinchin]\ \\
In a separable Hamiltonian system 
\begin{equation}\label{H}
H=\sum_{n=1}^N H_n(\vec{q}_n,\vec{p}_n),
\end{equation}
observables of the form
\begin{equation}\label{A}
A(\vec x)= \sum_{n=1}^N A_n(\vec{q}_n,\vec{p}_n), \;\;\; \text{ with } A_n=O(1),
\end{equation}
satisfy the relation
\begin{equation}\label{Khinchin}
\textnormal{Prob}\left( \frac{ \left| \overline{A}-\langle A \rangle \right|}{ \left| \langle A \rangle \right|}  \geqslant C_1 N^{-1/4} \right) \leqslant C_2 N^{-1/4}, \;\;\; \text{ with } C_1, C_2=O(1).
\end{equation}
\end{theorem}
  Essentially, the Khinchin theorem states the following: in a system of non-interacting particles, for the sum functions \eqref{A} (e.g. pressure, 
kinetic energy, single particle distribution), the set of points for which time and ensemble averages differ more than a given amount, which goes to zero as $N \to \infty$, has a measure which goes to zero as $N \to \infty$. 

Furthermore, under the assumptions \eqref{H} and \eqref{A} it is possible to prove that the relation \eqref{Khinchin} holds true even substituting $\overline{A} \to A$, i.e. physically relevant observables are in practice constant on a constant energy surface. This amounts to the stronger statement that statistical mechanics' approach, based on ensemble averages, is in fact independent of (mathematical) ergodicity. Its validity is purely a consequence of the large number of degrees of freedom of microscopic systems \cite{VULPIANI08}.

A weak aspect, from the physical point of view, of Khinchin's approach concerns
the no-interaction assumption.
 In contrast, an essential requisite for thermodynamic
behavior is the possibility of an exchange of energy among the particles.
Of course Khinchin noted the problem and argued that the actual Hamiltonian is
indeed only approximated by the separable Hamiltonian. The feeling of Khinchin
was that the interaction among the particles contributes very little to evaluating
the averages and for the majority of computations in statistical mechanics one can
neglect these terms.
The undesirable restriction to the separable structure of the Hamiltonian,
 was removed by Mazur and van der Linden \cite{MAZURLINDEN63}. They extended the
result to systems of particles interacting through a short-range potential, showing
that the intuition of Khinchin, that the interaction among the particles is of little
relevance, was basically correct. 
\\

\section{Non-integrable systems and the ergodic problem}
The issue of ergodicity is related to the problem of the existence of conserved quantities, also called first integrals, in Hamiltonian systems. Indeed, as integrable systems are non-ergodic, a natural question is whether non-integrability implies ergodicity.\\
First we recall the notion of integrable system:
\begin{definition}[Integrable system]\ \\
A system, described by the Hamiltonian $H(\vec{q},\vec{p})$ with $\vec{q},\vec{p} \in\mathbb{R}^N$, is called integrable if there exists a canonical transformation from the variables $(\vec{q},\vec{p})$ to the action-angle variables $(\vec I,\vec \theta)$, such that the new Hamiltonian takes the form $H=H_0(\vec I)$, i.e. it depends only on the action $\vec I$. 
\end{definition} 
The time evolution of an integrable system is given by:
\begin{align}\label{int}
\begin{dcases}
&\dot{\vec I}= -\frac{\partial H_0}{\partial \vec \theta}=\vec 0 \Longrightarrow \vec I(t)=\vec I(0), \\
&\dot{\vec \theta}=\frac{\partial H_0}{\partial \vec I}= \vec \omega \Longrightarrow \vec \theta(t)=\vec \theta(0)+\vec \omega(\vec{I}(0))t.
\end{dcases}
\end{align}
Hence, integrable systems have $N$ independent first integrals, as the action $\vec I$ is conserved and the motion evolves on $N$-dimensional tori $T_n$. Of course, from \eqref{int} it follows that $\overline{\vec \theta}= \vec \theta(0)$, which in general cannot coincide with $\langle \vec \theta \rangle$, integrable systems are evidently non-ergodic. Consequently, it is natural to wonder about the effect of a perturbation $\epsilon H_1(\vec I, \vec \theta)$ on $H_0$, and whether the perturbation allows for the existence of first integrals besides the energy. Clearly, that would maintain the perturbed system non-ergodic. This possibility is ruled out by Poincar\'{e} result \cite{POINCARE93}:
\begin{theorem}[Poincar\'{e}]\ \\
Consider a system described by the Hamiltonian 
\begin{equation}\label{hamiltonian}
H(\vec I,\vec \theta)= H_0(\vec I)+ \epsilon H_1(\vec I,\vec \theta).
\end{equation}
If $\epsilon=0$ the system has $N$ first integrals $\vec I$. Whereas, if  $0<\epsilon\ll1$, the system does not allow analytic first integrals, apart from energy.
\end{theorem}
 
The sketch of the proof goes as follows. We introduce the function 
\begin{equation}\label{F}
\vec{F}(t)= \vec{I} + \sum_m \epsilon^m \vec{F}_m(t) ,
\end{equation}
and require it to be a first integral, i.e.
\begin{equation}\label{poisson}
\frac{\diff \vec F}{\diff t}=\{ H, \vec F\}=\vec 0,
\end{equation} 
where $\{.\,,.\}$ denotes the Poisson bracket. Inserting \eqref{F} into \eqref{poisson} one obtains, for the lower meaningful order $O(\epsilon)$, the equation:
\begin{equation}\label{first}
\{H_0, \vec{F}_1\} + \{H_1, \vec{I}\}=\vec 0.
\end{equation}
If one expresses $H_1$ and $F_1$ as Fourier series in the angle vector $\vec \theta$:
\begin{align*}
&H_1(\vec{I}, \vec{\theta})= \sum_{\vec{k}} c_{\vec{k}}(\vec I) e^{- i\vec k \vec \cdot \vec \theta} , && \vec{F}_1(\vec{I}, \vec{\theta})= \sum_{\vec{k}} \vec{f}_{\vec{k}}(\vec I) e^{- i\vec k \vec \cdot \vec \theta},
\end{align*}
equation \eqref{first} yields after some algebra:
\begin{equation}\label{end}
\vec{f}_{\vec{k}}(\vec I)= \frac{c_{\vec{k}}(\vec{I}) \vec{k}\,}{\vec{\omega}_0(\vec I) \cdot \vec{k}},
\end{equation}
where $\vec{\omega}_0(\vec I)=\frac{\partial H_0}{\partial \vec I}$ is the unperturbed frequency vector. Clearly, since the denominator of \eqref{end} can be arbitrary small, $\vec{f}_{\vec{k}}$ may be divergent. Therefore, one concludes that first integrals, except for energy, cannot exist.\\
Nonetheless, it is not correct to infer from the Poincar\'{e} result, as it happened historically \cite{FERMI23}, that the Hamiltonian system \eqref{hamiltonian} becomes ergodic as soon as $\epsilon \neq 0$. A proof of the fact that non-integrability does not imply ergodicity is indeed given by the Kolmogorov-Arnold-Moser (KAM) theorem \cite{ARNOLD63, MOSER62}.

\begin{theorem}[KAM]\ \\
Given the Hamiltonian $H(\vec I,\vec \theta)= H_0(\vec I)+ \epsilon H_1(\vec I,\vec \theta)$, with $H_0$ sufficiently regular and $det\left| \frac{\partial^2 H_0(\vec I) }{\partial I_i \partial I_j}\right| \neq 0$, if $\epsilon \ll 1$, then, on the constant-energy surface, invariant tori $T_n^{*}$ survive in a region whose measure tends to 1 as $\epsilon \to 0$. The tori $T_n^*$ correspond to the tori $T_n$ of $H_0$ with deformations of order $O(\epsilon).$
\end{theorem}
  Therefore, the KAM theorem explains the lack of first integrals, revealed by the Poincar\'{e} result, with the local destruction of the (unperturbed) periodic orbits and, consequently, proves that small perturbations are in general not sufficient to provide ergodicity. In the next section we introduce a paradigmatic example of a nearly-integrable system, namely the Fermi-Pasta-Ulam chain, and qualitatively explain its properties in terms of the KAM theorem.

\subsection{The Fermi-Pasta-Ulam problem}

Fermi, Pasta and Ulam (FPU) investigated the time evolution of a chain of non-linear oscillators with pinned ends \cite{FERMI55}. Their primary interest was the thermalization time, i.e. the time needed by the system to lose memory of the initial conditions. In fact, since the non-linear forces render the system non-integrable, the physicists' community expected the system to display ergodic properties. This widespread belief was due to the conclusion, erroneously drawn by Fermi in view of the Poincar\'{e} result \cite{FERMI23}, according to which non-integrability should have implied ergodicity. The study of the FPU system first revealed the pitfall of this rationale, that later was better understood in the light of the KAM theorem.

\begin{figure}
\centering
\begin{subfigure}[b]{0.3\textwidth}
\includegraphics[width=1\textwidth]{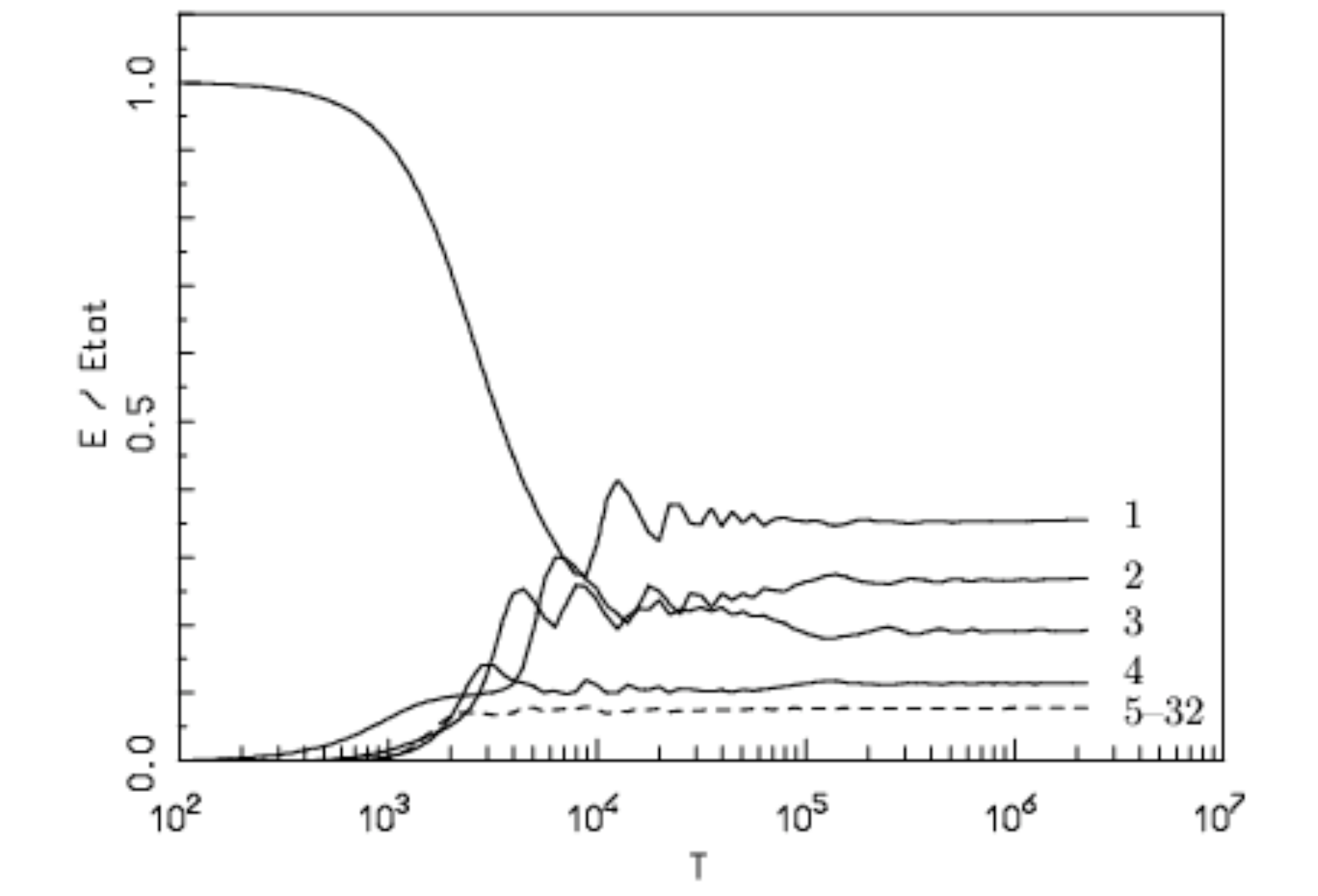}
\caption{$\xi<\xi^*$}
\label{fig:FPUa}
\end{subfigure}~
\begin{subfigure}[b]{0.3\textwidth}
\includegraphics[width=1\textwidth]{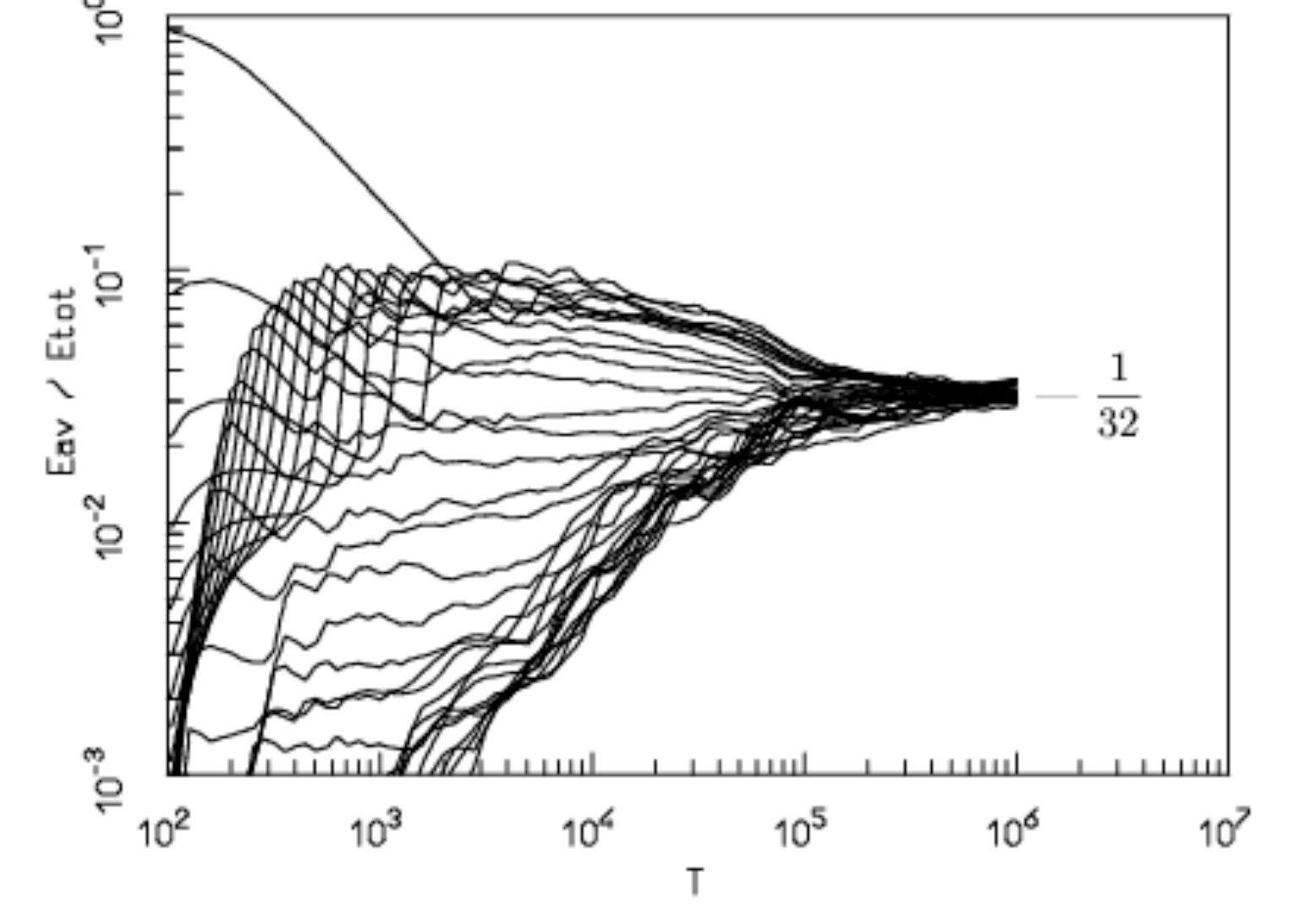}
\caption{$\xi>\xi^*$}
\label{fig:FPUb}
\end{subfigure}
\caption{Time-averaged energy per mode for the FPU systems with parameters 
$r=3, \,\epsilon=0.1$, $N=32$ and energy densities $E/N=0.07$ (\vref{fig:FPUa}) 
and $E/N=1.2$ (\vref{fig:FPUb}). In both cases the initial conditions are: $E_1(0)=E_0$ and $E_n(0)=0$ for $n\geq 2$. Before the results on the FPU system, it was expected that the time-averaged energy of each mode converges to the ergodic value of $E_0/N$. This is not true when the energy density is smaller than a characteristic value $\xi^*$ that depends on the size system and the perturbation intensity.}
\label{fig:FPU_energy_per_mode}
\end{figure}

The Hamiltonian considered in the FPU problem is:
\begin{equation}\label{eq:FPUT}
H=\sum_{n=0}^N\left[ \frac{p_n^2}{2m}+\frac{k}{2}\left( q_n-q_{n-1}\right)^2+\frac{\epsilon}{r}\left( q_n-q_{n-1}\right)^r \right],
\end{equation}
where $\epsilon$ is a perturbation parameter and $r=3$ or $r=4$. For $\epsilon=0$, the Hamiltonian \eqref{eq:FPUT} is integrable and can be written as 
\begin{align}\label{Nmodes}
H_0=\sum_{n=0}^N \left(\frac{\dot A_n^2}{2} + \frac{\omega_n^2}{2}A_n^2\right),
\end{align}
by exploiting the normal modes:
\begin{equation*}
A_n= \sqrt{\frac{2}{N+1}}\sum_{j} q_j \sin\left(\frac{ n \pi j}{N+1} \right).
\end{equation*}
Needless to say, the Hamiltonian $\eqref{Nmodes}$ describes a system of decoupled harmonic oscillators with angular frequencies $\omega_n$ expressed by
\begin{equation*}
\omega_n=\sqrt{\frac{4k}{m}}\sin\left( \frac{n \pi }{2(N+1)}\right).
\end{equation*}
For $\epsilon \ll1$ one can compute the equilibrium value of all the thermodynamic quantities as averages over a statistical ensemble. In particular, the energy per mode $\langle E_n \rangle$ is:
\begin{equation*}
\langle E_n\rangle= \frac{E_0}{N} +O(\epsilon),
\end{equation*}
which is nothing but the energy equipartition law.
Of course, with $\epsilon= 0$ the time average computed along a
(sufficiently long) trajectory, $\overline{E_n}$, cannot in general
coincide with $\langle E_n\rangle$ since different modes, being
uncoupled, maintain their initial energy. Instead, with $\epsilon > 0$
the expectation was to retrieve the equality $\langle
E_n\rangle=\overline{E_n}$ whatever the initial conditions $E_n(0)$,
even as weird as the one with excitation only in the first mode:
$E_1(0)=E_0$ and $E_n(0)=0$ for $n\geq 2$
\cite{CHI66,BNTTGLG2008}. However, numerical integration of
\eqref{eq:FPUT} for fixed $N$, $\epsilon>0$, and small energy density
$\xi=E/N$ proved this expectation wrong (see
fig.~\vref{fig:FPU_energy_per_mode}).
\\ Namely, there exists a threshold value of the perturbation parameter, $\epsilon_c$, function of $N$ and the initial conditions, which separates the following two regimes:
\begin{itemize}
\item if $\epsilon < \epsilon_c$, irrespective of the observation time, $\overline{E_n}$ depends on the initial conditions.
\item if $\epsilon > \epsilon_c$, $\overline{E_n}$ converges to the ergodic value $\langle E_n \rangle$.
 \end{itemize}
Notice that, when one deals with realistic working cases, the intensity of the perturbation is, usually, part of the model, i.e. not a control parameter. In these cases the control parameter can be the energy density, and a similar minimum value of $\xi$ exists.

These results can be seen as a verification of the KAM theorem  \cite{VULPIANI08}:
when  $\epsilon<\epsilon_c$  some of the invariant tori of the constant energy surface of the unperturbed system survive the perturbation and result in deformed tori. If the number of degrees of freedom is smaller than $N=2$, the tori separate regions of the phase space where, in principle, the system can show different behaviors. When  the number of degrees of freedom is bigger than 2, the complement set of the tori is connected, allowing for a \textit{diffusing} behavior among all the phase space, excluded the tori, called Arnold-diffusion \cite{ARNOLD64}.

\subsection{Is chaos necessary for the validity of the Statistical Mechanics?}

To proceed further we introduce the useful concept of Lyapunov exponent.
The Lyapunov exponent characterizes how much sensitive a system is with respect to initial conditions, quantifying the growth rate of the separation of trajectories initially at distance $\delta(0)\rightarrow0$.
\begin{definition}[Lyapunov exponent]\ \\
The first, or maximal, Lyapunov exponent of the dynamical system $(\Omega, S^t, \mu)$ is:
\begin{align}
\lambda = \lim_{t\rightarrow\infty}\lim_{|\vec{\delta}(0)| \rightarrow 0} \frac{1}{t} \ln \frac{|\vec{\delta}(t)|} {|\vec{\delta}(0)|}
\end{align}
where $\vec{\delta}(0)$ is a perturbation in the initial condition, namely $\vec{x} (0) \rightarrow \vec{x} '(0)=\vec{x} (0)+\vec{\delta}(0)$, and $\vec{\delta}(t)=S^t \vec{x} '(0)-S^t \vec{x} (0)$ measures the separation between trajectories. 
\end{definition}
 
The Lyapunov exponent allows to quantitatively measure the degree of instability of a system's evolution, introducing the precise notion of chaos: 
\begin{definition}[Chaos]\ \\
A dynamical system is called chaotic if its first Lyapunov exponent is positive.
\end{definition}
 
We note in passing that integrable systems have a null Lyapunov exponent and initial perturbations in the trajectories grow as polynomials.

Inspired by the discussion about the FPU problem, one is tempted to regard chaos as a necessary ingredient for the validity of statistical mechanics. Instead, we present here two examples, namely the FPU system itself and a model of coupled rotators, to show that chaos is neither sufficient nor necessary to ergodic behavior.\\ 
	In the FPU system, at fixed value of the perturbation parameter $\epsilon$, the Lyapunov exponent is positive whatever the value of the energy density $\xi$ is. However, one observes thermalization only when the energy density is larger than a threshold value $\xi >\xi_c$. This would suggest that chaos,  despite not being sufficient, is at least necessary to ergodicity.\\ Conversely, a ready counter-example is provided by the XY model, where a non-chaotic regime is accompanied by ergodicity breakdown. The model consists of a chain of $N$ coupled rotors of angular momentum $m$ and orientations $\theta_n \in [0,2\pi)$:
\begin{align}\label{x Y}
H=\sum_{n=1}^N \left[\frac{p^2}{2m}+\epsilon \left(1-\cos(\theta_n-\theta_{n-1})\right)\right]
\end{align}
There are two opposite limits when the system becomes integrable: at
very small energy density $\xi \to 0$, when the interaction is
approximately harmonic, and at very large energy density $\xi \to
\infty$, when the rotors are almost independent. The Lyapunov exponent
$\lambda$ approaches zero in these two limits, but has a maximum in
between. Notwithstanding the presence
of chaos for sufficiently large $\xi$, it was observed
\cite{LIVI87,LIVI85} that the ensemble averaged values of the specific
heat $C_v$ does not match the results of numerical simulations of \eqref{x Y},
i.e. with  the time average of the  fluctuations of energy   in subsystems.
On the contrary in the other integrable limit (i.e. small energy), although
the system is not chaotic (or very weakly chaotic),
the specific heat $C_v$ computed with the time average of the fluctuations of 
energy in subsystems,
is in perfect agreement  with the results of the canonical ensemble. 

The different behavior of $C_v$ in the two near-integrable regimes of low and high
temperature can be understood as follows. For the FPU system and for the low temperature
rotators the ``natural''  variables are the normal modes, which, even in
a statistical analysis, are able to show regular behavior and where the energy of the
system is resident. However, even if the normal modes are almost decoupled, when
observing the energy of a subsystem, identified by some set of ``local'' variables
$\{q_j , p_j \}$, non-negligible fluctuations of the ``local'' energy can be seen. On the
other hand, for the chain of rotators at large energy the normal modes, i.e. the
carriers of the energy, are the``local'' variables $\{q_j , p_j \}$ themselves, and therefore
the fluctuations of the local energy are strongly depressed, as is the exchange of
energy among the subsystems.
\\

\section{Mixing}
If a deterministic dynamical system is described by the flow $\dot{\vec{x }}=\vec F(\vec{x })$, the associated probability density evolves in time according to the Liouville equation:
\begin{equation}\label{liouville}
\partial_t\rho(\vec x ,t)+ \vec{F}(\vec x) \cdot \partial_{\vec{x }} \rho(\vec x ,t) =0.
\end{equation}
The invariant density, if it exists, is the density $\rho(\vec x )$ that satisfies the equation $ \vec{F} \cdot \partial_{\vec{x }} \rho(\vec x ) =0$. As we have seen, if the dynamical behavior is periodic, the system keeps memory of the initial condition $\rho(\vec x , 0)$ and no relaxation to the invariant density takes place. The sufficient condition that ensure the invariant density to be eventually reached is mixing \cite{ARNOLD68}:
\begin{definition}[Mixing]\ \\
A dynamical system $(\Omega, S^{t}, \mu)$ is called mixing if, for all measurable sets $D, E \subset \Omega$, one has
\begin{equation}\label{mixing}
\lim_{t\to \infty} \mu(D \cap S^{-t} E)=\mu(D)\mu(E).
\end{equation}
\end{definition}
 
In words, the fraction of points starting from $D$ and ending up in $E$ after a long time is the product of the measures of $D$ and $E$, whatever their position in $\Omega$. A consequence of relation \eqref{mixing} is that generic functions of $\vec x $ become uncorrelated in the long time limit:
\begin{equation}
\lim_{t \to \infty} \langle A(\vec x (t)) B(\vec x (0))\rangle = \langle A(\vec x ) \rangle \langle B(\vec x )\rangle,
\end{equation}
where the average is on the invariant measure. Of course, the mixing condition is stronger that ergodicity, i.e. mixing systems are ergodic of necessity  \cite{ARNOLD68}.

\section{Chaos and coarse-graining}
 
In this section we discuss more throughly the role of chaos in dynamical system and in particular its repercussions on mesoscopic level of descriptions, i.e. when coarse-graining approximations are performed. This is of extreme importance for very many physical cases, where microscopic dynamics are in general not amenable to exact analytical studies and experimental observations due, respectively, to the inherent complexity and the finite resolution of the measurements. With this aim, we introduce, via an information theory approach, and exploit the concept of entropy of a dynamics system, trying to address the following questions: 
\begin{itemize}
\item what informations about the system can entropy provide and how is it affected by different coarse-graining approximations?
\item how much do we know of the  coarse-grained system from the microscopic Lyapunov exponent? Are there general results in this direction?
\end{itemize}
These questions are tightly linked, but for clarity will be discussed separately. 

\subsection{Information theory entropy}

Consider a set of symbols $\{s_1,...,s_T\}$ as the outcome of a certain process, e.g. integers belonging to the set $s_i \in \{1,...,M\}$ emitted by a source. In order to measure the information content of the $N$-word ensemble one associates to the $N$-word $W_N=(s_1,...,s_N)$ the block entropy:
\begin{definition}[Block entropy]\ \\
Given an ensemble of words $\{W_N\}$, the block entropy is:
\begin{equation}\label{entropy}
H_N\equiv-\sum_{\{W_N\}} P(W_N) \ln P(W_N),
\end{equation}
where $P(W_N)=P(s_1,...,s_N)$ is the occurrence probability of the word $W_N$.
\end{definition}
 
In the following the occurrence probability is assumed stationary.
The block entropy allows one to define a central quantity linking information theorem and statistical mechanics, namely, the Shannon entropy \cite{SHANNON48}:
\begin{definition}[Shannon entropy]\ \\
The Shannon entropy is defined as the entropy per symbol:
\begin{equation}\label{shannon}
h_{\text{Sh}}\equiv \lim_{N\to\infty} \frac{H_N}{N}=\lim_{N\to \infty} \left(H_N - H_{N-1}\right),
\end{equation}
\end{definition}
 
The Shannon entropy quantifies the asymptotic uncertainty about the further emission of a symbol $s$ in a very long sequence. The connection between the Shannon entropy and the thermodynamic entropy is given by the Shannon-McMillan theorem  \cite{KHINCHIN57}:
\begin{theorem}[Shannon-McMillan]\ \\
The ensemble of $N$-words can be divided into two classes, $\Omega_1(N)$ and $\Omega_0(N)$, such that, when $N\to \infty$:
\begin{align*}
&\sum_{W_N \in \Omega_1(N)} P(W_N)\longrightarrow 1, && \sum_{C_N \in \Omega_0(N)} P(C_N) \longrightarrow 0,
\end{align*}
and all the words $W_N \in \Omega_1(N)$ have the same probability $P(W_N) \sim \exp(-N h_{\textnormal{Sh}})$.
\end{theorem}
 
Put in other words, the theorem states that the occurrence probability $P(W_N)$ assigned to a (very long) sequence will be close to $\exp(-N h_{\textnormal{Sh}})$ and independent of the specific sequence. Therefore, the Shannon entropy counts the number of effectively observed sequences $\mathcal{N}(\Omega_1(N))$, i.e. $N h_{\textnormal{Sh}} \sim \ln \mathcal{N}(\Omega_1(N))$, exactly as the  thermodynamic entropy, $S= k_{\text{B}} \ln W$, counts the number of microscopic configurations $W$ corresponding to a macroscopic state of a physical system.  \\ 

\subsection{Entropy in coarse-grained dynamical systems}

Let us now apply the notions introduced in the previous paragraph to a deterministic dynamical system.
 First, we perform a partition $\mathcal{A}$ of the phase space $\Omega$, with $\{s\}_{\mathcal{A}}$ numbering the partition cells, and define a sampling time $\tau$. Then, we construct the word $W_N=(s(i),...,s(i+N-1))$, where $s(i)$ labels the cell visited by the trajectory $\vec{x}(t)$ at time $i\tau$. If the system is ergodic,  observing the frequencies of the words $W_N$, we can construct the block entropies $H_N(\mathcal{A})$ and the Shannon entropies $h_{\text{Sh}}(\mathcal{A})$ as defined, respectively, in \eqref{entropy} and \eqref{shannon}. In order to get a quantity independent of the chosen partition one introduces the Kolmogorov-Sinai (K-S) entropy \cite{KOLMOGOROV58, SINAI59}:
\begin{definition}[Kolmogorov-Sinai entropy] \ \\
Given a set of partitions $\{\mathcal{A}\}$ of the phase space of the dynamical system $(\Omega,S^t,\mu)$, the Kolmogorov-Sinai entropy is defined as:
\begin{align}\label{KSentr}
h_{\textnormal{KS}}=\sup_\mathcal{A} \,h_{\textnormal{Sh}}(\mathcal{A}).
\end{align}
\end{definition}
 
Thanks to the Shannon-McMillan theorem, a precise physical meaning can be assigned to $h_{\text{KS}}$. Namely, called $N(\epsilon, t)$ the number of sizeably probable trajectories originating form a cell of edge $\epsilon \to 0$, the K-S entropy expresses their exponential growth rate in the long-time limit:
\begin{equation}\label{KSrate}
h_{\text{KS}}t \sim \ln N(\epsilon, t).
\end{equation}
There is  an important relation between $h_{\text{KS}}$ and the Lyapunov exponent. Indeed, the K-S entropy of typical Hamiltonian systems can be expressed in terms of the Lyapunov exponents  \cite{OTT93}:
\begin{align}
h_{\text{KS}}=\sum_{\lambda_n>0} \lambda_n
\end{align}
In very low-dimensional chaotic systems, where only one Lyapunov exponent is generally positive, the K-S entropy is simply $h_{\text{KS}}=\lambda_1$.

The Kolmogorov-Sinai entropy of a system gives a measure of the amount of information per unit of time needed to reproduce a trajectory with arbitrary precision [\cite{VULPIANI08}, sec 2.3.4]. Nevertheless, in reality finite resolution makes impossible to compute $h_{\textnormal{KS}}$. Therefore, we have to content ourselves with measuring the amount of information needed to reproduce trajectories with finite accuracy  $\epsilon$. To this end, one introduces partitions $\mathcal{A}_{\epsilon}$ based on cells of size $\epsilon$ that define a Shannon entropy $h_\text{Sh}(\mathcal{A}_\epsilon, \tau)$, where the $\tau$ dependence indicates the fine time resolution. Making the latter quantity independent of the partition, one introduces the $\epsilon$-entropy  \cite{GASPARD93}:
\begin{definition}[$\epsilon$-entropy]\ \\
The $\epsilon$-entropy is the infimum over all partitions $\mathcal{A}_\epsilon$ whose cell diameter is smaller that $\epsilon$:
\begin{align}
h(\epsilon,\tau)=\inf_{\mathcal{A}_\epsilon : \textnormal{diam}(\mathcal{A}_\epsilon)<\epsilon}h_{\textnormal{Sh}}(\mathcal{A}_\epsilon,\tau)
\end{align}
\end{definition}
It is easy to convince oneself that in the limit of infinite accuracy, i.e. when $\epsilon \to 0$, the K-S entropy is recovered for deterministic systems:
\begin{align}\label{zerolimit}
h_{\textnormal{KS}} =\lim_{\epsilon\rightarrow0} h(\epsilon)
\end{align}
Furthermore the $\epsilon$-entropy is also a proper definition of entropy in stochastic systems where the Kolmogorov-Sinai entropy is ill defined due to the absence of the supremum required by definition \eqref{KSentr} \cite{GASPARD93}. \\
An important result for the $\epsilon$-entropy in stochastic dynamical system is given by the theorem \cite{KOLMOGOROV56}:
\begin{theorem}[Kolmogorov] \ \\
Given a Gaussian stochastic process $x (t)$ whose mean square displacement behaves as:
\begin{equation}
\langle \left(x (t+\Delta t)-x (t)\right)^2\rangle \propto \Delta t^{2 \alpha}
\end{equation}
with $0\leq\alpha<1$.
The $\epsilon$-entropy is:
\begin{equation}
\lim_{\tau \rightarrow 0}h(\epsilon,\tau)\propto \,\epsilon^{-\frac{1}{\alpha}}\label{KOLMOGOROVDIFF}
\end{equation}
\end{theorem}

\subsection{Chaos and diffusion}
We are now in the position to discuss the questions we asked at the beginning of this section. As first example consider the  following discrete-time evolution law \cite{VULPIANI08,STROGATZ01}:
\begin{align}\label{map}
x _{t+1}=[x _t]+f(x _t-[x _t])
\end{align}
where $[x]$ means the integer  part of $x $, and $f(y)$ is defined as:
\begin{equation*}
f(y)=
\begin{cases}
(2+\alpha)y  &\text{ if } y\in[0,1/2)\\
(2+\alpha)y - (1+\alpha)  &\text{ if } y\in[1/2,1]
\end{cases}
\end{equation*}
where $\alpha>0$. The maximum Lyapunov exponent is computed as:
\begin{align}
\lambda_1 = \ln \bigg|\frac{df}{dy}\bigg|=\ln |2+\alpha| >0,
\end{align}
which shows that the map is chaotic for every value of $\alpha$. In addition, for long observation times the map \eqref{map} displays a diffusive behavior \cite{GEISEL84} characterized by the diffusion coefficient $D$:
\begin{align}
\langle (x _t-x _0)^2 \rangle \approx 2Dt,
\end{align}
where $\langle.\rangle$ denotes an average over the initial conditions.
In view of the relations \eqref{zerolimit} and \eqref{KOLMOGOROVDIFF}, one obtains the following behavior of the $\epsilon$-entropy:
\begin{align}
h(\epsilon) \sim &\, h_{\text{KS}}=\lambda_1 && \text{if } \epsilon \ll1, \label{small}\\
h(\epsilon) \sim &\, \frac{D}{\epsilon^2} && \text{if } \epsilon \gg1. \label{large}
\end{align}
The latter relations show that:
\begin{itemize}
\item when $\epsilon\ll1$  the $\epsilon$-entropy reduces to the Kolmogorov-Sinai one. Since the first Lyapunov exponent determines the value of the entropy, the dynamics is characterized by the fastest time scale and the microscopic details.
\item for $\epsilon\gg 1$ the last relation suggests the existence of a de-correlation time so that, at large scales, the microscopic details become unimportant.
\end{itemize}

At first glance, one may be tempted to think that deterministic diffusion
can appear with positive Lyapunov exponents only. 
For example, think of a free particle moving
through hard obstacles. If the obstacles are convex and positioned on
a lattice (Lorentz-Sinai billiard), the Lyapunov exponent is positive
and the long-time behavior is diffusive \cite{SINAI79}. Conversely, if
the obstacles are randomly distributed polygons (wind-tree Ehrenfest
model), the Lyapunov exponent is zero, but the long-time behavior is
still diffusive.  Therefore, one has to conclude that chaotic behavior
does not imply the existence of Arnold-diffusion, nor vice versa. We
can understand that the Lyapunov exponent discussed so far is given by
the microscopic details of the dynamics and thus is unable to
characterize the macroscopic behavior.\\

Up to this point we have seen that, in low dimensional systems:
\begin{itemize}
\item we can associate an entropy to every chaotic system. Its value depends on the coarse-graining approximation.
\item when the coarse-graining scale is at ``atomic'' scales, the entropy is given by the first Lyapunov exponent.
\item when the coarse-graining scale is coarser, the entropy of a chaotic system is given by an effective ``diffusion'' coefficient. But we have no information about the Lyapunov exponents of the macroscopic observables.
\end{itemize}

\subsection{Lyapunov exponent and coarse-graining}
\begin{figure}[h!t]
\centering
%\begin{subfigure}[b]{0.3\textwidth}
\includegraphics[width=.5\textwidth]{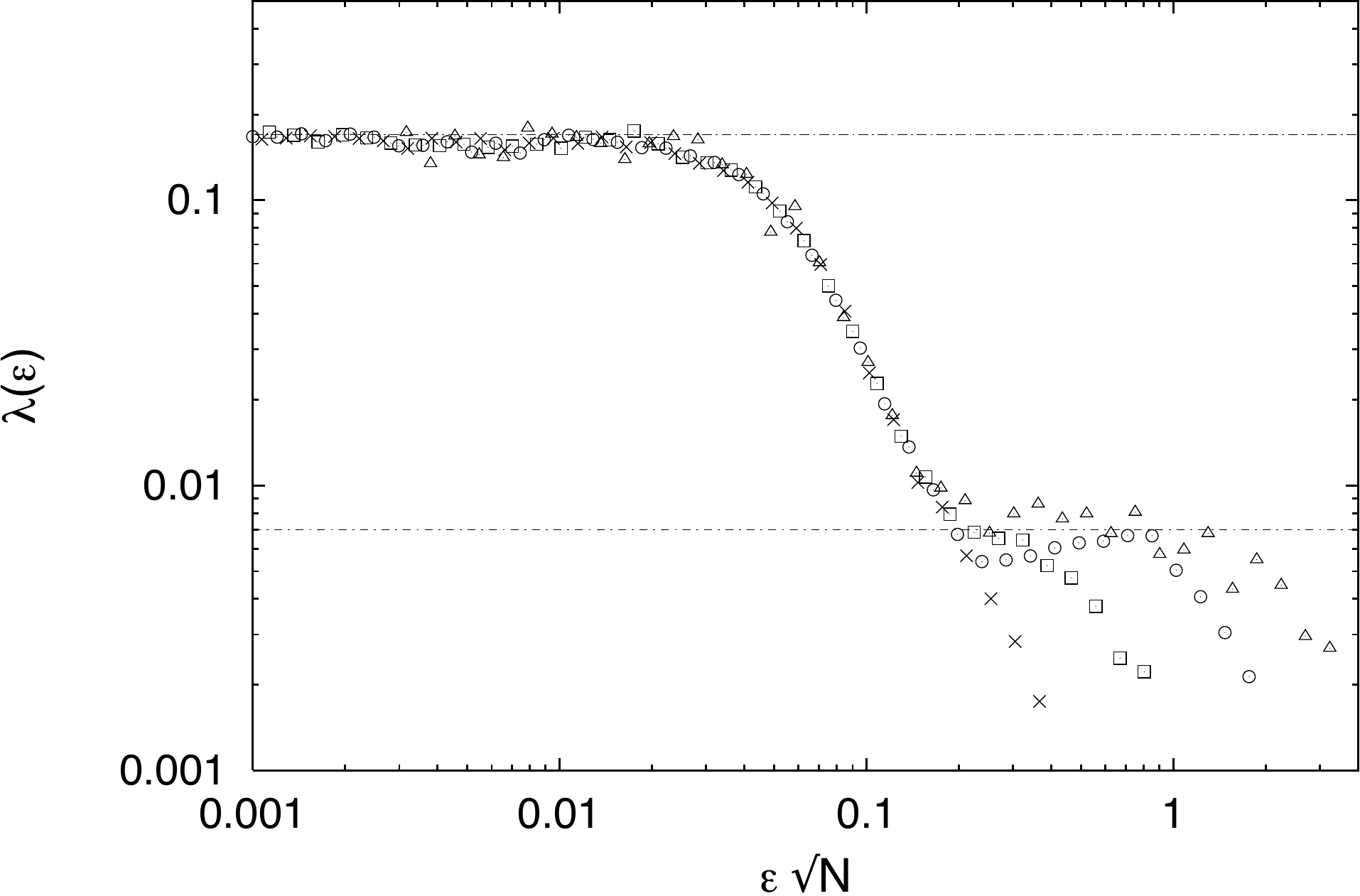}
%\caption{Lyapunov exponent for the FPU system}
%\label{fig:FPUc}
%\end{subfigure}~
%\begin{subfigure}[b]{0.3\textwidth}
%\includegraphics{dumbo}
%\caption{Lyapunov exponent for the x Y model}
%\label{fig:x Ya}
%\end{subfigure}
\caption{The Finite Size  Lyapunov Exponent $\lambda(\epsilon)$ for  the map \eqref{map2}
with $c=0.3, \, a=1.7, \, N=10^4, \, 10^5, \, 10^6, \, 10^7$,
 as function of the coarse-graining scale $\epsilon/\sqrt{N}$. The Lyapunov exponent for the microscopic scale $(\epsilon\rightarrow0)$ reaches the plateau-value $\lambda_1$ that is larger than the Lyapunov exponent for the macroscopic scale $\lambda_M$. This is a ``universal'' property that does not depend on the specific map. The macroscopic Lyapunov exponent saturates at about $\epsilon\approx1/\sqrt{N}$.}
\label{fig:lambdavseps}
\end{figure}

We discuss now the relation between the Lyapunov exponent of a microscopic description of a dynamical system, and the Lyapunov exponent of the coarse-grained one. 
As  example we discuss  to the globally coupled map of $N$ interacting particles:
\begin{align}\label{map2}
x _n(t+1)=(1-c)f(x _n(t))+\frac{c}{N}\sum_{n'\neq n}^N g_c(x _{n'}(t)),
\end{align}
with $g_c(x)=(1/2-|x-1/2|)$.
The above system can be considered  as a picture of a gas of $N$ interacting particles. A macroscopic observable associated to the microscopic dynamics \eqref{map2} can be the centre of mass position:
\begin{align}
M(t)=\frac{1}{N}\sum_n x _n=\langle x _n \rangle +o\left(\frac{1}{\sqrt{N}}\right).
\end{align}
We call $\lambda_1$  and $\lambda_M$ the first Lyapunov exponents of the observables $x _n(t)$ and $M(t)$, respectively. 
The behavior of the  Lyapunov exponent for different coarse-graining approximations 
was studied numerically with the so called Finite Size Lyapunov Exponent $\lambda(\epsilon)$
(which is based on the same idea of the $\epsilon$ entropy
 \cite{SHIABATA98,CENCINI99}) yielding the following limit behaviors:
\begin{align}
\lambda(\epsilon)\simeq  &\lambda_M\leq\lambda_1 &\text{when } \epsilon\ll\frac{1}{\sqrt{N}} \label{micro} \\
\lambda(\epsilon)\simeq &\lambda_1&\text{when } \epsilon\gg\frac{1}{\sqrt{N}} \label{macro}
\end{align}
The behavior  of $\lambda(\epsilon)$ has a characteristic coarse-graining threshold $\epsilon\sim \frac{1}{\sqrt{N}}$ separating the microscopic Lyapunov exponent \eqref{micro} from the macroscopic one \eqref{macro}. 
To be more precise, for the value of $\lambda(\epsilon)$ we observe two plateaus: $\lambda(\epsilon<\epsilon_1)\approx\lambda_1$ and $\lambda(\epsilon>\epsilon_2)\approx\lambda_M$. The values of $\epsilon_1$ and $\epsilon_2$ are of the order of  $1/\sqrt{N}$ and generically $\lambda_1\geq\lambda_M$ (see fig.~\vref{fig:lambdavseps}).
\\
%\FloatBarrier
\section{Some final general remarks}
Let us conclude  with some remarks on ergodicity and chaos with respect
to the foundation of statistical mechanics.
First we note that the ergodic approach can be seen as a natural way to introduce
probabilistic concepts in a deterministic context. It seems to us that the ergodic
theory provides support for the frequentistic interpretation of probability in the
foundation of statistical mechanics. The other way (which is not in disagreement
with the point of view of Boltzmann) to introduce probability is to assume an
amount of uncertainty in the initial conditions. This approach is due to Maxwell
who considers that there are {\it a great many systems the properties of which are the
same, and that each of these is set in motion with a different set of values for the
coordinates and momenta}.
Since one is forced to deal with a unique system (although with many degrees
of freedom) it seems natural to assume that the purpose of statistical mechanics,
for equilibrium phenomena, is to calculate time averages according to the temporal
evolution of the system. 

Therefore the ensemble theory should be seen only as a
practical mathematical tool and the ergodic theory (or a ``weak'' version, such as
that of Khinchin and Mazur and van der Linden) is an unavoidable step. Of course
there is no complete consensus on this; for example Jaynes's opposite opinion is
that ergodicity is simply not relevant for the Gibbs method.

The ergodicity is, at the same time, an extremely demanding property (i.e. the
time and phase averages must be equal for almost all the initial conditions), and not
very conclusive at a physical level (because of the average over an infinite time).
On the other hand, in the quasi-integrable limit the analytical results (KAM theorem)
 give only qualitative indications and do not allow for quantitative
aspects. Therefore it is not possible to avoid detailed numerical investigations.
Let us note that in the numerical computations based on  the molecular dynamics
 one basically assumes  that ergodicity somehow holds (although
not in a strict mathematical sense) for the physically relevant observables.

There are also opposing answers to the question of whether the systems which are
described by statistical mechanics must have a large number of degrees of freedom,
and it is possible to find eminent scientists with opposite opinions. For instance
according to Grad {\it the single feature which distinguishes statistical
mechanics is the large number of degrees of freedom}. One can read rather similar
sentences in the well known textbook of Landau and Lifshitz. In contrast Gibbs
believed that {\it the laws of statistical mechanics apply to conservative systems
of any number of degrees of freedom, and are exact}.
Extended simulations on high-dimensional Hamiltonian systems show in a clear
way that chaos is not necessarily a fundamental ingredient for the validity of equilibrium
statistical mechanics: the na\"ive idea that chaos implies good statistical properties
is inconsistent. Indeed sometimes, even in the absence of chaos (in agreement
with Khinchin's ideas), one can have good agreement between the time averages
and their values predicted by equilibrium statistical mechanics.

\bibliography{VulpianiNotes}

\begin{thebibliography}{10}
\expandafter\ifx\csname url\endcsname\relax
  \def\url#1{\texttt{#1}}\fi
\expandafter\ifx\csname urlprefix\endcsname\relax\def\urlprefix{URL }\fi
\expandafter\ifx\csname href\endcsname\relax
  \def\href#1#2{#2} \def\path#1{#1}\fi

\bibitem{NAGEL1979}
E.~Nagel, The structure of science, Hackett Publishing Company, 1979.

\bibitem{ARNOLD68}
V.~I. Arnold, A.~Avez, Ergodic problems of classical mechanics, Benjamin New
  York, 1968.

\bibitem{BIRKHOFF31}
G.~D. Birkhoff, Proof of the ergodic theorem, PNAS 17 (1931) 656.

\bibitem{KAC59}
M.~Kac, Probability and related topics in physical sciences, Vol.~1, AMS
  Bookstore, 1959.

\bibitem{KHINCHIN57}
A.~I. Khinchin, Mathematical foundations of statistical mechanics, New York
  Dover, 1957.

\bibitem{VULPIANI08}
P.~Castiglione, M.~Falcioni, A.~Lesne, A.~Vulpiani, Chaos and Coarse Graining
  in Statistical Mechanics, Cambridge University Press, 2008.

\bibitem{MAZURLINDEN63}
P.~Mazur, J.~van~der Linden, Asymptotic form of the structure function for real
  systems, J. Math. Phys. 4 (1963) 271.

\bibitem{POINCARE93}
H.~Poincar{\'e}, Les m{\'e}thodes nouvelles de la m{\'e}canique c{\'e}leste,
  Vol.~2, Paris: Gauthier-Villars, 1893.

\bibitem{FERMI23}
E.~Fermi, Dimostrazione che in generale un sistema meccanico normale \'{e}
  quasi-ergodico, Il Nuovo Cimento (1911-1923) 25 (1923) 267.

\bibitem{ARNOLD63}
V.~I. Arnold, Proof of a theorem of a. n. kolmogorov on the preservation of
  conditionally-periodic motions under small perturbations of the hamiltonian,
  Russ. Math. Surv. 18 (1963) 9.

\bibitem{MOSER62}
J.~K. Moser, On invariant curves of area-preserving mapping of an annulus,
  Nachr. Akad. Wiss. G\"{o}ttingen Math. Phys. Kl. 2 (1963) 1.

\bibitem{FERMI55}
E.~Fermi, J.~Pasta, S.~Ulam, Studies of nonlinear problems, Tech. rep., I, Los
  Alamos Scientific Laboratory Report No. LA-1940 (1955).

\bibitem{CHI66}
F.~M. Izrailev, B.~V. Chirikov, Statistical properties of a nonlinear string,
  Sov. Phys. Dokl. 166 (1966) 57.

\bibitem{BNTTGLG2008}
G.~Benettin, A.~Carati, L.~Galgani, A.~Giorgilli, The fermi—pasta—ulam
  problem and the metastability perspective, in: G.~Gallavotti (Ed.), The
  Fermi-Pasta-Ulam Problem, Vol. 728 of Lecture Notes in Physics, Springer
  Berlin Heidelberg, 2008.

\bibitem{ARNOLD64}
V.~I. Arnold, Instability of dynamical systems with many degrees of freedom,
  Dokl. Akad. Nauk SSSR 156 (1964) 9.

\bibitem{LIVI87}
R.~Livi, M.~Pettini, S.~Ruffo, A.~Vulpiani, Chaotic behavior in nonlinear
  hamiltonian systems and equilibrium statistical mechanics, J. Stat. Phys. 48
  (1987) 539.

\bibitem{LIVI85}
R.~Livi, M.~Pettini, S.~Ruffo, M.~Sparpaglione, A.~Vulpiani, Equipartition
  threshold in nonlinear large hamiltonian systems: the fermi–pasta–ulam
  model, Phys. Rev. A 31 (1985) 1039.

\bibitem{SHANNON48}
C.~E. Shannon, A mathematical theory of communication, Bell Syst. Tech. J. 27
  (1948) 379--423, 623--656.

\bibitem{KOLMOGOROV58}
A.~N. Kolmogorov, A new metric invariant of transient dynamical systems and
  automorphisms in lebesgue spaces, in: Dokl. Akad. Nauk. SSSR, Vol. 119, 1958,
  p. 861.

\bibitem{SINAI59}
Y.~G. Sinai, On the concept of entropy of a dynamical system, in: Dokl. Akad.
  Nauk. SSSR, Vol. 124, 1959, pp. 768--771.

\bibitem{OTT93}
E.~Ott, Chaos in dynamical systems, Cambridge University Press, 1993.

\bibitem{GASPARD93}
P.~Gaspard, X.~J. Wang, Noise, chaos, and ($\epsilon$ , $\tau$)-entropy per
  unit time, Phys. Rep. 235 (1984) 291.

\bibitem{KOLMOGOROV56}
A.~N. Kolmogorov, On the shannon theory of information transmission in the case
  of continuous signals, IRE Trans. Inf. Theory 1 (1956) 102.

\bibitem{STROGATZ01}
S.~H. Strogatz, Nonlinear Dynamics And Chaos: With Applications To Physics,
  Biology, Chemistry, And Engineering, 1st Edition, Westview Press, 2001.

\bibitem{GEISEL84}
T.~Geisel, J.~Nierwetberg, Intermittent diffusion: a chaotic scenario in
  unbounded systems, Phys. Rev. A 29 (1984) 2305.

\bibitem{SINAI79}
Y.~G. Sinai, Ergodic properties of the lorentz gas, Funct. Anal. Appl. 357
  (1979) 143.

\bibitem{SHIABATA98}
T.~Shibata, K.~Kaneko, Collective chaos, Phys. Rev. Lett. 81 (1998) 4116.

\bibitem{CENCINI99}
M.~Cencini, M.~Falcioni, D.~Vergni, A.~Vulpiani, Macroscopic chaos in globally
  coupled maps, Physica D 130 (1999) 58.

\end{thebibliography}
\bibliographystyle{elsarticle-num}

\end{document}